\documentclass[notitlepage,superscriptaddress,prl,reprint]{revtex4-1}
\usepackage{url,graphicx,braket,mathdots,physics,xfrac, amssymb, amsmath}
\begin{document}
%
%
\title{Rotational spectroscopy of a single molecular ion at sub part-per-trillion resolution}
\author{Alejandra L. Collopy}
\affiliation{National Institute of Standards and Technology, Boulder, CO 80305, USA}
\author{Julian Schmidt}
\affiliation{National Institute of Standards and Technology, Boulder, CO 80305, USA and Department of Physics, University of Colorado, Boulder, CO 80309, USA}
\author{Dietrich Leibfried}
\affiliation{National Institute of Standards and Technology, Boulder, CO 80305, USA and Department of Physics, University of Colorado, Boulder, CO 80309, USA}
\author{David R. Leibrandt}
\affiliation{National Institute of Standards and Technology, Boulder, CO 80305, USA and Department of Physics, University of Colorado, Boulder, CO 80309, USA}
\author{Chin-Wen Chou}
\affiliation{National Institute of Standards and Technology, Boulder, CO 80305, USA and Department of Physics, University of Colorado, Boulder, CO 80309, USA}
%
\begin{abstract}
We use quantum-logic spectroscopy (QLS) and interrogate rotational transitions of a single CaH$^+$ ion with a highly coherent frequency comb, achieving a fractional statistical uncertainty for a transition line center of 4$\times$10$^{-13}$. We also improve the resolution in measurement of the Stark effect due to the radio-frequency (rf) electric field experienced by a molecular ion in an rf Paul trap, which we characterize and model. This allows us to determine the electric dipole moment of CaH$^+$ by systematically displacing the ion to sample different known rf electric fields and measuring the resultant shifts in transition frequency.
\end{abstract}

\maketitle
\begin{figure}[h]
	\includegraphics[width=.85\columnwidth]{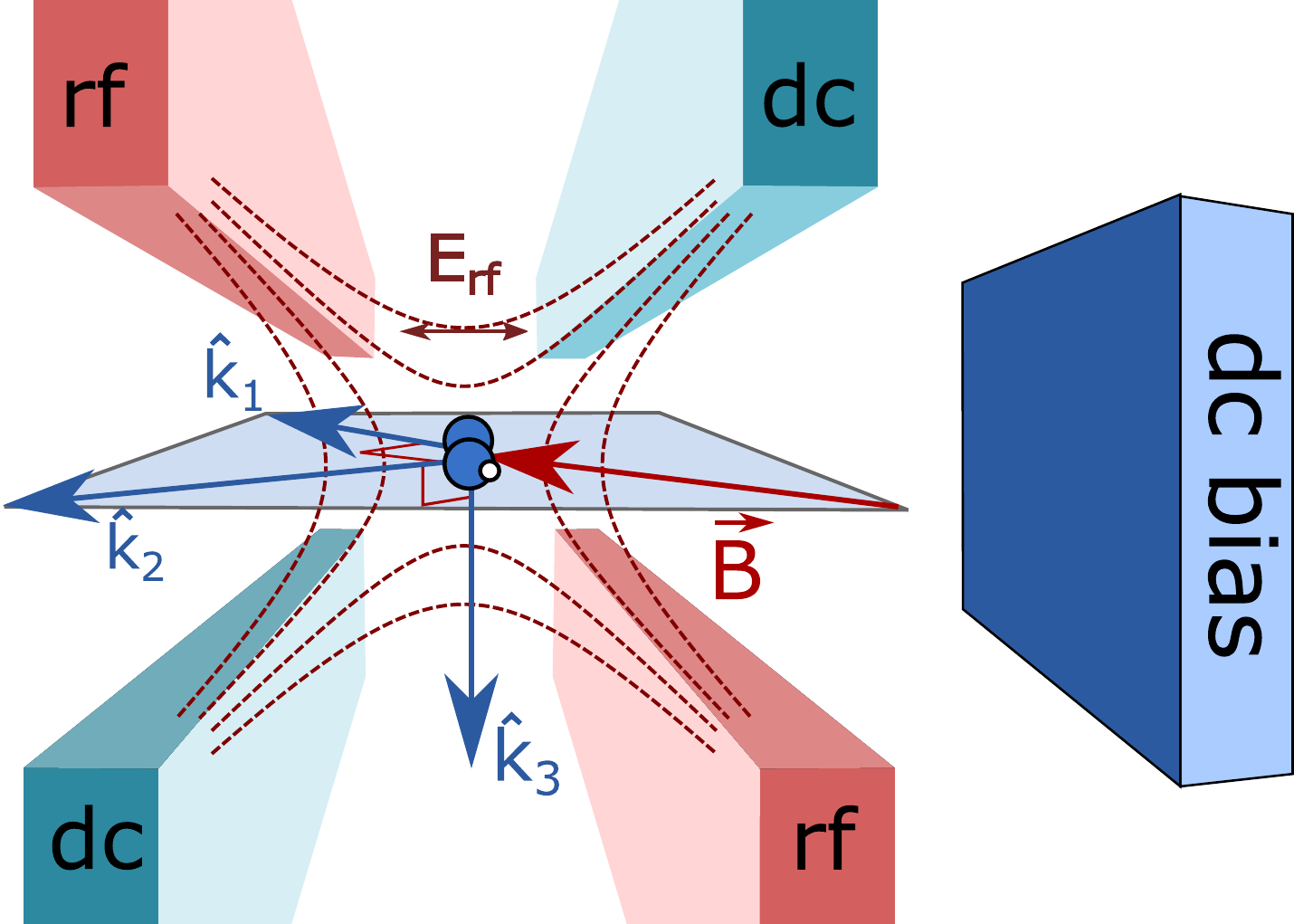}
	\caption{An ideal linear rf trap forms an electric quadrupole field with a null along the trap axis. Ions positioned off of the null experience an rf electric field. Due to the electrode geometry of the trap used in our experiments, the rf-electric field along the axis has a finite minimal amplitude.  By adjusting the voltages applied to the dc electrodes and the more distant dc bias electrode, the ions can be intentionally shifted from this minimum to sample different rf electric field magnitudes and directions. Three 729~nm laser beams are used to measure micro-motion along approximately mutually orthogonal directions, $\hat{k}_{1}$ parallel to the magnetic quantization axis $\vec{B}$, $\hat{k}_{2}$, and $\hat{k}_{3}$.
	}	
	\label{fig:trapdiagram}
\end{figure}
\begin{figure}[h]
	\includegraphics[width=1.05\columnwidth]{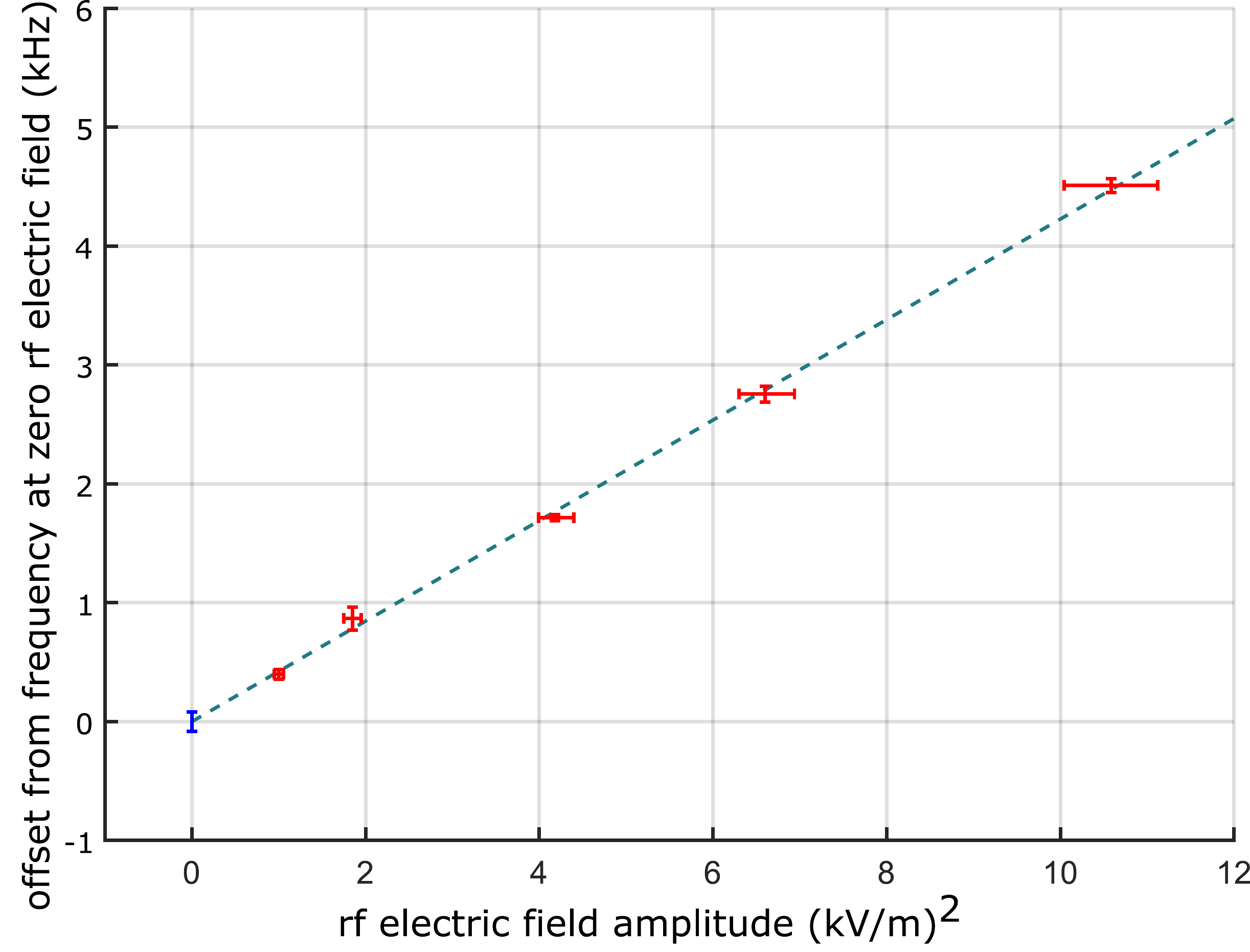}
	\caption{Red markers show the measured shifts in frequencies of the  $\ket{2,-3/2,-}$ to $\ket{0,-1/2,-}$ transition as a function of the squared rf electric field amplitude at the molecular ion location. Spectroscopic probe times varied from 2.5 to 20.5~ms for different data points. The blue marker shows the rf electric field-free transition frequency (95 $\%$ confidence interval [-82, 80]~Hz) obtained through extrapolation to zero field with uncertainties determined by parametric bootstrapping. The dashed green line is a linear fit to the data. Its slope can be used to determine the permanent dipole moment of the ion.}
	\label{fig:transvselectric} 
\end{figure}

Significant efforts are currently devoted to achieve precision spectroscopy and quantum state control of molecules. A much larger variety of molecules exists, compared to atoms, and they possess richer structures that promise entirely different functionalities and increased suitability for certain tasks, for example greater sensitivities for various tests of fundamental physics \cite{Kajita2014,Andreev2018,Cairncross2017,Grasdijk2021}. High internal state coherence and the potential for conversion of quantum information across many decades in frequency also make molecules attractive for quantum information processing \cite{Demille2002,Hudson2018,Albert2020,Najafian2020a,Lin2020}. Despite impressive progress in recent years, quantum state preparation, detection, and control of molecules remains more difficult than for atoms  \cite{Wolf2016,Sinhal2020,Burchesky2021,Fan2021, Li2021}.\\ \indent
Quantum-logic spectroscopy (QLS)~\cite{Schmidt2005} shows great promise and versatility for the study of charged particles in general and molecular ions in particular. It relies on an atomic ``logic'' ion species for performing sympathetic cooling of the joint translational motion and for state readout, and enables quantum-state preparation, manipulation, and spectroscopy of charged particles (``spectroscopy'' ions) that are difficult to control otherwise~\cite{Chou2017,Chou2019,Micke2020}. All lasers addressing the molecular ion in our experiments drive far-detuned stimulated two-photon Raman transitions that do not rely on a particular level structure of a molecule. This, along with the fact that sympathetic cooling of translational degrees of freedom and quantum-logic readout can also be performed with few requirements on the details of the molecular structure, makes QLS available to a wide variety of ion species.\\ \indent
To explore new applications of molecules, it is also critical to achieve high resolution and to account for minute systematic effects that become relevant at this unprecedented level of precision. In particular, spins and the relative motion of nuclei add degrees of freedom and coupling mechanisms to the environment that are not present in atoms. For example, the ac Stark shift suffered by neutral polar molecules when trapped in optical lattices or tweezers needs to be carefully controlled to retain long coherence times~\cite{Kotochigova2010, Gregory2021, Burchesky2021}, and understanding of trap shifts due to electric and magnetic fields are critical for high-precision molecular spectroscopy in ion traps \cite{Cairncross2017, Alighanbari2020, Patra2020}.\\ \indent
In this Letter, we improve the resolution of two-photon stimulated Raman spectroscopy on the rotational states of the CaH$^+$ ion, which enables precise investigation of the effects of the radio-frequency (rf) electric field in an rf trap on rotational transition frequencies. We perform spectroscopy at sub-part-per-trillion precision by driving a rotational transition at $\sim$2 terahertz with a highly coherent frequency comb. Rf electric fields at the molecular ion position can lead to coupling and level shifts of order part-per-billion that can be precisely measured and used to determine the permanent electric dipole moment of the molecular ion when combined with independent in-situ sensing of the rf electric field with the Ca$^+$ logic ion.\\ \indent
Trapped ions in rf traps can undergo micro-motion~\cite{Berkeland1998,Keller2015}  -- position-dependent driven motion at the angular frequency $\Omega_{\rm rf}= 2 \pi f_{\rm rf}$ of the rf drive that provides confinement of the ions. Some ion traps are specifically designed to have high symmetry to minimize the rf electic field at the trap center, but even then the time dilation shift and Stark shift due to the rf electric field need to be accurately measured and controlled for precision experiments \cite{Brewer2019}. Various methods have been developed to characterize micro-motion and the electric field that drives it \cite{Berkeland1998,Keller2015}. In dipolar molecular ions, the rf electric field $\vec{E}_{\rm rf}$ can interact with the permanent electric dipole moment $\vec{d}$ of the molecule via the interaction Hamiltonian
\begin{equation}
H = -\vec{d}\cdot\vec{E}_{\rm rf}\cos(\Omega_{\rm rf} t),    
\end{equation}
and thereby affect rotational states $\ket{J,M}$ with principal quantum number $J$ and projection quantum number $M$. For linear molecules, the perturbation in $H$ first order in $\vec{E}_{\rm rf}$ vanishes~\cite{Townes1975,supplement}. The second order yields the matrix element between rotational states $\ket{J,M}$ and $\ket{J',M'}$ for the quadratic Stark effect
\begin{multline}\label{Eq:MatEle}
\bra{J',M'} H^{(2)} \ket{J,M}=\\ \frac{\cos^2(\Omega_{\rm rf} t)}{4 h}\sum_{i} \frac{\bra{J',M'}\vec{d}\cdot\vec{E}_{\rm rf}\ket{i}\bra{i}\vec{d}\cdot\vec{E}_{\rm rf}\ket{J,M}}{f_{\rm rf}-\nu_{J,i}} \\ +\frac{\bra{J',M'}\vec{d}\cdot\vec{E}_{\rm rf}\ket{i}\bra{i}\vec{d}\cdot\vec{E}_{\rm rf}\ket{J,M}}{-f_{\rm rf}-\nu_{J,i}},
\end{multline}
where the sum runs over all intermediate states with non-zero matrix elements, namely $\ket{i}=\ket{J_i,M_i}$, $0\leq J_i=J\pm 1$, $M_i =M +\{-1,0,1\}$ and $|M_i| \leq J_i$. The frequency $\nu_{J,i} = (E_i-E_{J})/h $ is proportional to the difference between rotational state energies $E_J$ and $E_{i}$ and $h$ is the Planck constant. Time-averaging over one period of the trap and neglecting the term oscillating at $2 \Omega_{\rm rf}$ yields $\cos^2(\Omega_{\rm rf} t)\approx 1/2$. Setting $E_{\rm rf}=|\vec{E}_{\rm rf}|$ reveals that the matrix elements are of order  $C_J=d^2 E_{\rm rf}^2/[4 h B_R 2(J+1)]$, small compared to the rotational energy $E_{\rm rot}\approx h B_R J(J+1)$ for typical dipole moments and amplitudes of the rf-electric field. For example, the rotational constant of CaH$^+$ is  $B_R=142.5017779(17)$ GHz~\cite{Chou2019} while the matrix elements are $\leq h$(1 kHz) for typical fields. Only matrix elements of nearly degenerate states with $J=J'$ are non-negligible and $f_{\rm rf}$ ($\sim$ 86~MHz in this work) can be neglected compared to $\nu_{J,i}$ in Eq.(\ref{Eq:MatEle}). Higher order rotational constants~\cite{Chou2019} can be taken into account for $E_J$, but will contribute below 10$^{-3}$ relative to the lowest order proportional to $B_R$ for the states we explored experimentally. In this approximation the matrix element for $J=0$ is
\begin{equation}\label{Eq:GSshift}
H^{(2)}_{0,0}=-\frac{d^2 E_{\rm rf}^2}{12 h B_R}\text{.}
\end{equation}
For fixed $J>0$, $|M_J|\leq J$ and the rf field decomposed into spherical components (see \cite{supplement}) $\vec{E}_{\rm rf}= E_{\rm rf}(\epsilon_{-1} \vec{n}_{\sigma^-}+\epsilon_0 \vec{n}_\pi+\epsilon_{1} \vec{n}_{\sigma^+})$, the non-zero matrix elements $\bra{J,M'} H^{(2)} \ket{J,M}=H^{(2)}_{M,M'}$ are
\begin{widetext}
\begin{eqnarray}\label{Eq:MatElExp}
H^{(2)}_{M,M}&=&C_J \frac{J(J+1)-3 M^2}{J(2J-1)(2J+3)}[2 \epsilon_0^2-(\epsilon_{-1}^2+\epsilon_1^2)] \nonumber \\
H^{(2)}_{M,M+1}&=&C_J \frac{3(2 M+1)\sqrt{J+J^2-M(M+1)}}{\sqrt{2}\,J(2J-1)(2J+3)}\epsilon_0(\epsilon_{1}-\epsilon_{-1})\nonumber\\
H^{(2)}_{M,M+2}&=&C_J \frac{3\sqrt{(J-M-1)(J-M)(J+M+1)(J+M+2)}}{J(2J-1)(2J+3)}\epsilon_{-1} \epsilon_1.
\end{eqnarray}
\end{widetext}
For an rf-field along the quantization axis, $\epsilon_0=1$, $\epsilon_{-1}=\epsilon_1=0$, $H^{(2)}_{M,M}$ agrees with the result in \cite{Townes1975}, Chapter 10, Eq. (10.8) when setting the average square of the rf-electric field equal to $E_{\rm rf}^2/2$ .\\ \indent
We evaluate the effect of the trap rf electric field by driving transitions within and between rotational manifolds of the CaH$^+$ ion subjected to a variety of rf electric field configurations, improving the uncertainty of the rf electric field-free frequency of a rotational transition by more than an order of magnitude \cite{Chou2019}. Additionally, characterizing the effect of varying the magnitude of trap rf electric field at the molecular ion enables measurement of the molecular electric dipole moment of the CaH$^+$ ion for the first time. The electric dipole moment of other suitable molecular ions could be determined through the same procedure.\\ \indent
The steps to prepare a pure quantum state of the molecule have been described before~\cite{Chou2019}. In brief, Doppler cooling, electromagnetically-induced-transparency cooling, and resolved sideband cooling of the Ca$^+$-CaH$^+$ ion crystal is performed on the Ca$^+$ ion. Several of the motional modes shared by both ions are cooled to the ground state, including the axial out-of-phase mode, which is used for QLS. A series of motional sideband pulses on the molecule involving the QLS mode followed by ground state cooling on the atomic ion (``pumping cycles'')~\cite{Chou2017} are used to drive the molecule from a mixture of states determined by the thermal environment towards certain target states in rotational manifolds with $1\leq J \leq 6$ within the $^1\Sigma$ vibronic ground state. Each target state has an associated ``signature transition'' with a unique frequency. Sequentially driving motion-adding sidebands on the signature transitions can add a quantum of motion to the QLS mode if the molecule was in the corresponding target state. The quantum of motion can be detected by driving a motion-subtracting sideband on the Ca$^+$ ion and reading out its final internal state. Ideally, if the readout indicates a successful motion subtracting transition on Ca$^+$, the molecule is projected into the final state of its preceding motion adding transition. The sequence prepares the molecule in a known pure quantum state and heralds it non-destructively.\\ \indent
Following preparation, further spectroscopic pulses can be driven which may change the molecular internal state. If the spectroscopic pulse changes the state of the molecule, departure from the initial state can be detected in a similar way as arrival during preparation, signaled by a negative result. Repeated cycles allow for accumulation of statistics. Stimulated Raman two-photon transitions that change $J$ by $\pm 2$ are driven by a Ti:Sapph frequency comb centered at $\sim$800~nm, as described in more detail in \cite{Chou2019}. Compared to that work, the comb has been improved. Following~\cite{Bartels2009}, a small fraction ($\approx$ 40 mW) is split off the frequency comb output and spectrally broadened with a nonlinear fiber to facilitate stabilizing the repetition rate to a narrow-linewidth continuous-wave laser as an optical frequency reference ~\cite{supplement}. The optical frequency reference improves the comb coherence, which enables longer probe times and higher spectral resolution of transitions in our setup. The unbroadened main portion of the frequency comb output, after proper power control, is used to perform rotational spectroscopy as described in~\cite{Chou2019}.
\begin{figure}[h]
	\includegraphics[width=\columnwidth]{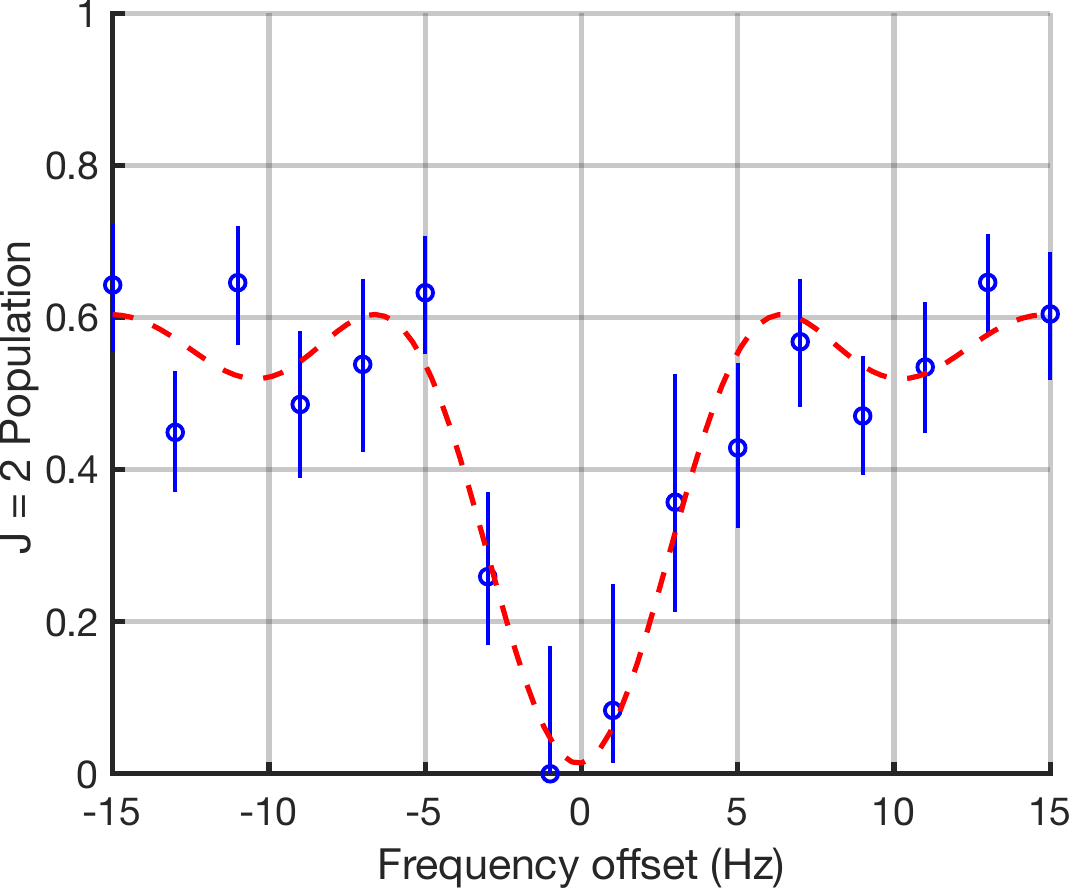}
	\caption{Population in the initial $\ket{J=2, m=-5/2,-}$ rotational state vs. the Raman difference frequency offset to the center of the line model (dashed red line)  of the $\ket{2,-5/2,-} \rightarrow \ket{4,-7/2,-}$ rotational transition at approximately 2 THz. Error bars indicate a 68 $\%$ confidence interval. Dashed line shows the best fit of a Rabi lineshape with free parameters of center frequency offset, remaining population, contrast, and Rabi rate, with pulse duration fixed. The fit yields a 0.8~Hz statistical uncertainty (95 $\%$ confidence) of the offset of the data points to the model line center. The 128 ms probe time is a substantial fraction of the black-body lifetime of the initial state, resulting in reduced contrast of the line.} 
	\label{fig:combfreqscan} 
\end{figure}
Because the magnetic moment of the hydrogen nucleus and the rotational magnetic moment couple together, energy eigenstates of CaH$^+$ are superpositions of the product states $\ket{J,M}\ket{\frac{1}{2},\pm\frac{1}{2}}$, where the second ket represents the magnitude and projection of the proton spin on the quantization axis. We use the notation $\ket{J,m,\xi}$~\cite{Chou2017,Chou2019}, where $m=M\pm 1/2$ denotes the joint projection of rotational angular momentum and proton spin along the quantization axis, which remains a good quantum number and $\xi=\{+,-\}$ labels one of the two eigenstates that share the same $J$ and $m$, except for $\ket{J,m=\pm(J+1/2),\pm} = \ket{J,\pm J}\ket{1/2,\pm 1/2}$. After preparing the molecule in $\ket{2,-5/2,-}$, a 128~ms square envelope pulse train of the Raman beams derived from the frequency comb is applied to the molecular ion, followed by an attempt to detect the initial molecular state. Figure \ref{fig:combfreqscan} depicts the result of scanning the frequency difference between the Raman beams over the resonance of the transition to $\ket{J'=4,-7/2,-}$ at $\sim$2~THz (line Q $\approx 3.3\times 10^{11}$). A fit to a Rabi lineshape for a square spectroscopy pulse \cite{Wineland1998} yields a statistical uncertainty of the line center of $\sim$0.8~Hz, corresponding to a fractional statistical uncertainty of $4\times10^{-13}$. This compares favorably with the resolution of rotational spectroscopy on molecular ions in other experiments~\cite{Patra2020,Alighanbari2020}.\\ \indent
To characterize the effect of the rf electric field on rotational transitions in the molecular ion, we vary and measure the direction and amplitude of the field at the molecular ion. As schematically shown in Figure \ref{fig:trapdiagram}, we utilize nearby electrodes to manipulate the position of the ions and thereby the rf electric field amplitude and direction experienced by the molecular ion. Imperfections of the electrode geometry of the trap used in our experiments results in a nonzero minimal rf-field amplitude of approximately 1~kV/m.
We can characterize the rf electric field by exchanging the positions of the atom and the molecule and using the atom to probe the rf-electric field amplitude and direction by measuring micro-motion sideband-to-carrier Rabi-rate ratios of the narrow 729~nm quadrupole transition \cite{Keller2015} (also see supplemental information~\cite{supplement}). Three 729~nm laser beams are directed along the magnetic quantization axis as well as two directions orthogonal to that axis and each other respectively. The experimentally determined Rabi-rate ratios can, with some assumptions, be inverted to yield three orthogonal components of the rf electric field amplitude~\cite{supplement}.\\ \indent
Rotational states of CaH$^+$ with the same $J$ can have energy splittings smaller than 1 kHz in the  approximately 0.36 mT static magnetic field in our experiment. The rf electric field can cause significant mixing of near-degenerate electric field-free eigenstates of the molecule ~\cite{Chou2017} and modeling of transitions becomes quite involved~\cite{supplement}. A relatively simple case where spin, rotation, and electric field decouple arises for the extreme $\ket{J,\pm(J + 1/2),\pm}$ states when rf electric field and static magnetic field are parallel~\cite{supplement}, but technical constraints in our setup prevent changing the rf electric field magnitude in this arrangement. 
Because the J=0 rotational level is the most sensitive to the rf electric field, we measure the frequency of the  $\sim$855~GHz transition $\ket{2,-3/2,-}$ to $\ket{0,-1/2,-}$ as a function of the electric field amplitude due to the rf drive, as shown in Figure \ref{fig:transvselectric}. Vertical error bars represent the 95$\%$ confidence interval of the line center when fitting transition line shapes similar to the one shown in Fig. \ref{fig:combfreqscan}, while horizontal errors are set by uncertainty in determining the electric field at the position of the molecule~\cite{supplement}. Fitting of the slope and intercept is performed by parametric bootstrapping  \cite{supplement}. The rf electric field-free transition frequency can be found from the intercept of this fit with a 95$\%$ confidence interval of [-82, 80] Hz. This uncertainty is more than one order of magnitude smaller than previous measurements \cite{Chou2019}.\\ \indent
The dependence of the rotational transition frequency on the rf electric field can be used to determine the electric dipole moment of the molecule. In particular, we can bound the shift of the initial state $\ket{2,-3/2,-}$, which has a particularly small dependence on the rf electric field, by modeling the effect of the field on the eigenstates of the molecule and their energies~\cite{supplement}. We find that $\ket{2,-3/2,-}$ shifts by no more than 50 Hz for any of the rf-amplitudes used in the experiments and account for this by adding 50 Hz in quadrature to the frequency uncertainties shown in Fig. \ref{fig:transvselectric}. In this approximation, the frequency shifts can be attributed to the state $\ket{0,-1/2,-}$, described by Eq.(\ref{Eq:GSshift}), and only depend on the amplitude of the rf electric field experienced by the molecule. The slope of the fit function shown in Fig.~\ref{fig:transvselectric} yields a value of $d= 17.81 \pm{0.63} \times 10^{-30}$\,C\,m (5.34 $\pm~{0.19}$~Debye)
for the permanent electric dipole moment of CaH$^+$, with the 95 $\%$ confidence interval found by parametric bootstrapping, in agreement with previous theory values of 5.310~Debye \cite{Abe_2010} and 5.13(2)~Debye \cite{Plessow}. To the best of our knowledge, this is the first experimental determination of the permanent dipole moment of a molecular ion using this method.\\ \indent
In future work, we can improve the uncertainty of field-free transition frequency measurements due to the rf electric field by using a more symmetric trap with smaller minimal rf electric field amplitude at the position of the molecule. Uncertainties in determining micro-motion could be improved by better control of the 729~nm beam directions, or by using different methods~\cite{Keller2015}. State preparation can be more efficient by using the frequency comb and/or a microwave source addressing the $J''=0$ to $J'=1$ transition to prepare molecules in a target rotational state. Our approach is suitable for molecular spectroscopy and electric dipole measurements on a variety of other molecular species, given that the molecule of interest is not too disparate in charge-to-mass ratio from that of the logic ion. We demonstrate versatile control techniques based on QLS that are stepping stones towards interrogating increasingly complex molecules, such as polyatomics, with the potential to distinguish isotopomers, isomers and possibly even the chirality of molecular ions, which may be particularly advantageous for tests of fundamental physics \cite{Patterson2018, Hutzler2020}.\\ \indent
In conclusion, we demonstrate sub-part-per-trillion resolution rotational spectroscopy of CaH$^+$ using methods that are suitable for a wide range of molecular ion species, including some of interest for tests of fundamental physics. We characterize the Stark effect due to the rf electric field experienced by the molecular ion and account for its effect on the measured transition frequencies, improving the systematic uncertainty of an rf electric field-free rotational transition frequency in the molecular ion by an order of magnitude from our previous work \cite{Chou2019}. In addition, we determine the electric dipole moment of the CaH$^+$ molecular ion, demonstrating the use of the rf field in an ion trap for measuring molecular properties.
\begin{acknowledgements}
We thank D. Hume for alerting us to the potential effects of the rf trap drive on transition frequencies, and L. Sinclair and Y. Liu for careful reading and their suggestions on the manuscript. This work was partially supported by the U.S. Army Research Office (grant no. W911NF-19-1-0172). During this work A.L.C. was supported by a National Research Council postdoctoral fellowship. J.S. gratefully acknowledges support from the  Alexander von Humboldt Foundation.
\end{acknowledgements}
\bibliographystyle{apsrev4-1}
\bibliography{micromotion}
\end{document}


\title{Supplementary Material for  ``Rotational spectroscopy of a single molecular ion at sub part-per-trillion resolution"}

\author{Alejandra L. Collopy}
\affiliation{National Institute of Standards and Technology, Boulder, CO 80305, USA}
\author{Julian Schmidt}
\affiliation{National Institute of Standards and Technology, Boulder, CO 80305, USA and Department of Physics, University of Colorado, Boulder, CO 80309, USA}
\author{Dietrich Leibfried}
\affiliation{National Institute of Standards and Technology, Boulder, CO 80305, USA and Department of Physics, University of Colorado, Boulder, CO 80309, USA}
\author{David R. Leibrandt}
\affiliation{National Institute of Standards and Technology, Boulder, CO 80305, USA and Department of Physics, University of Colorado, Boulder, CO 80309, USA}
\author{Chin-Wen Chou}
\affiliation{National Institute of Standards and Technology, Boulder, CO 80305, USA and Department of Physics, University of Colorado, Boulder, CO 80309, USA}

\maketitle

\vspace{200mm}
\pagebreak
\section{level diagram}

\begin{figure}[htb!]
	\includegraphics[width=0.82\columnwidth]{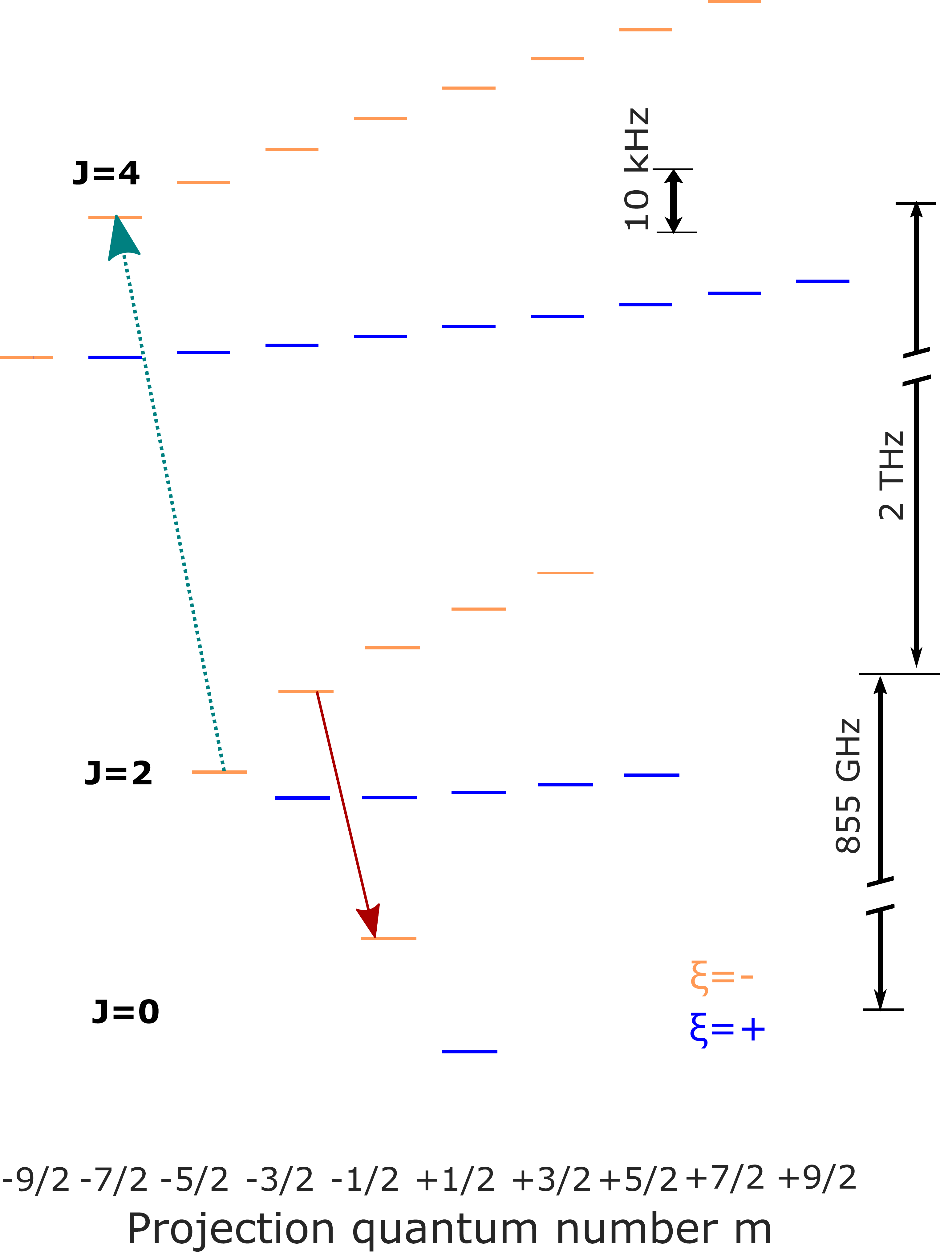}
	\caption{CaH$^+$ rotational states used in this work. Within a rotational manifold state splittings are on the order of tens of kHz, shown approximately to scale, while the rotational constant $B_R$ is $\sim$143 GHz, so differences between different $J$-states are not to scale. Frequency comb driven Raman transitions from state $\ket{J''=2,-5/2,-}$ to $\ket{J'=4,-7/2,-}$ (green dashed arrow) were used for our highest resolution data shown in main text Figure 1, while determination of the molecular dipole moment in main text Figure 3 was done on the $\ket{2,-3/2,-}$ to $\ket{0,-1/2,-}$ transition (red arrow). States with $\xi$=$-$ are shown as orange horizontal lines, while states with $\xi$=$+$ are blue.} 
	\label{fig:leveldiagram} 
\end{figure}

\section{duration of experiments}
%
\noindent Collecting sufficient data for the determination of the rf-electric field at the position of the molecular ion (see below)
takes approximately 1 minute for each data point shown in Fig. 3 of the main text. Subsequent determination of the frequency of the $\ket{2,-3/2,-} \leftrightarrow \ket{0,-1/2-}$ transition took 30-90 minutes for each point. Drifts of the rf-electric field at the position of the ion while data taking is completed for one measurement point are negligible compared to the other sources of uncertainty. Data acquisition for the $\ket{2,-5/2,-} \rightarrow \ket{4,-7/2,-}$ transition shown in Fig. 1 of the main text took about 80 minutes of experimental run time.
\section{rf electric field measurement}

\noindent After adjusting the trap electrode voltages to change the position of the ion pair, we wish to measure the rf electric field experienced by the molecular ion. To do so, we use three laser beams entering the vacuum chamber on nominally orthogonal axes (see Figure 2 
in the main text for a depiction of the configuration of electrodes and laser beams). All are 729~nm laser beams addressing the Ca$^+$ $S_{1/2}\rightarrow D_{5/2}$ quadrupole transition. As described in \cite{Keller2015}, the rf electric field component along a laser beam direction can be written as
\begin{equation}
E_{\rm rf} = \frac{m_{\rm ion} \Omega_{\rm rf}^2}{kQ}\beta 
\end{equation}
where $m_{\rm ion}$ is the ion mass, $Q$ the ion charge, $k$ the wavenumber, and $\Omega_{\rm rf}$ the trap drive frequency. The modulation index $\beta$ can be related (for $\beta \ll 1 $) to measured Rabi rates by
\begin{equation}
\frac{\Omega_{\pm 1}}{\Omega_0} = \frac{J_1(\beta)}{J_0(\beta)}\approx \frac{\beta}{2}
\end{equation}
where $J_n$ is the \emph{n}th order Bessel function of the first kind, $\Omega_0$ is the carrier Rabi rate, and $\Omega_{\pm 1}$ is the micromotion sideband Rabi rate.

An intensity stabilization servo was implemented for each 729~nm beam to ensure that the  $\pi$-time measurements of the carrier and micromotion sideband transitions can be directly translated to Rabi rate ratios, which then yield the rf electric field's component along each 729-nm beam's $k$ vector from Eq. (1). 

The carrier to micromotion sideband Rabi rate ratio measurements were performed on the Ca$^+$ ion. During the spectroscopy experiments the ion chain axial order is inverted (Ca$^+$ and CaH$^+$ axial positions exchanged). Because of the differing masses of the Ca$^+$ and CaH$^+$ ions, they experience different radial confinement and so will sample different radial rf electric fields for a given shim voltage that pushes the ions off the trap axis. An axial rf electric field gradient would be orders of magnitude smaller than that seen in the radial directions, and so we assume the axial component of the rf electric field is the same for the CaH$^+$ as for the Ca$^+$ when in the opposite ion order.

 The mode orientation of the trap radial modes with respect to the 729~nm laser beams was determined by measuring the mode frequencies as well as the 729~nm sideband $\pi$  times for  in-phase modes.  This information allows determination of trap parameters that represent our ion trap potential \cite{Wineland1998}. A numerical minimization of the potential energy of two ions in a trap with these parameters in addition to a static electric field shows that the CaH$^+$ ion is displaced from the RF null by a factor of  1.059 (1.024) of the Ca$^+$ ion along the lower (higher) frequency radial mode direction for a given rf-electric field gradient. Due to the quadrupole nature of the rf-potential, this indicates that the electric field along the radial mode direction experienced by the CaH$^+$ ion is also scaled up by the same value compared to that experienced by the Ca$^+$ ion. 
 
 This radial rf electric field correction factor is accounted for in the rest of our analyses by decomposing the electric field experienced by the Ca$^+$ ion into the coordinate system of the orthogonal trap mode directions, and applying the appropriate radial rf electric field scaling.
 
\section{rf electric field configurations for J'=2 to J''=0 measurements}

\noindent For the lowest rf electric field magnitude data point in Figure 3 of the main text we used voltages on dc electrodes to push the molecular ion to a position with minimized radial rf electric field, so that it experiences only an rf electric field along the trap axis. Since the magnetic quantization field is at 45 degrees to the trap axis, this electric field couples rotational states with $M \neq M'$, but has the lowest available rf-electric field amplitude we were able to find in our trap. For data points with higher rf electric field amplitude, we positioned the molecular ion so that it experiences an electric field that is orthogonal to the magnetic quantization axis, and can be expressed with $\epsilon_0 \simeq 0$ and $|\epsilon_{-1}|\simeq|\epsilon_1|$.

\section{RF magnetic field induced effects}

\noindent It has been noted \cite{Gan2018} that magnetic fields oscillating at the trap drive frequency can be erroneously attributed to the micromotion sideband modulation index that we attribute to the rf electric field. In our system we investigated the possibility of this effect affecting our measurement by looking at micromotion sideband-to-carrier Rabi rate ratios on various initial and final sublevels for the Ca$^+$ $S_{1/2}\rightarrow D_{5/2}$ transitions with differing sensitivities to magnetic field (g factors) under two different micromotion compensation voltage conditions. We determine that in our system this effect is not large enough to systematically shift our electric field measurements beyond their stated uncertainty, but this depends on details of the ion trap and is worth checking in other experimental apparatuses.

\section{Parametric bootstrapping}

\noindent In order to fit the $J=2 \rightarrow J=0$ transition data versus rf electric field magnitude we used parametric bootstrapping to account for uncertainties in transition frequencies and rf-electric field amplitude. Here we give a more detailed description of the data analysis procedure. 

We begin by assigning uncertainties for each data point in both the horizontal (rf electric field magnitude) and vertical (transition frequency) directions. Repeatedly measuring $\pi$-pulse times of carrier and micromotion sidebands in order to determine the rf electric field strength, we find that the standard deviation of measured $\pi$ times is less than .009 of the mean. We thus conservatively assign an uncertainty of 0.01 of the mean to the $\pi$-time measurements. Angular deviations of the 729~nm laser beams with respect to their nominally orthogonal axes also contribute uncertainty, which we estimate to be a standard deviation of 1 degree from normal for most directions, though one directional component of the vertically aligned beam can be better constrained to 0.38 degrees based on trap geometry. Uncertainty of the transition frequency for a given data point is determined by the uncertainty of the fit to the spectral line (see Fig. 1 of the main text for an example). 

We then bootstrap synthetic data points by drawing angular deviations, $\pi$ times, and transition frequencies from normal distributions with the aforementioned standard deviations. For a given set of 5 different rf electric fields, we bootstrap each using the same  drawn parameters affecting the uncertainty in the electric field. By using the measured electric field values along each beam direction and the sampled angular deviations, we determine the electric field that would yield that measurement, and use that value for the electric field for the bootstrapped data point. Then a linear fit is performed with these 5 bootstrapped points, yielding an intercept and a slope. By repeating this process $\sim$800 times, we then find a distribution of fit intercepts and slopes, allowing us to assign 95$\%$ confidence intervals to each.

\section{Rotational eigenstates in the presence of magnetic and electric fields}
\noindent To lowest order in the principal quantum number $J$, rotational angular momentum eigenstates $\ket{J,M}$ of the angular momentum operator $\vec{J}$ of a linear molecule with 
$$
\vec{J}\,^2\ket{J,M}=J(J+1)\hbar^2\ket{J,M},~ J\geq 0 ~{\rm and}~J_z \ket{J,M}=M\hbar\ket{J,M},~-J \leq M\leq J,
$$
where $\hbar=h/(2 \pi)$, have energies 
$$
E_J = h B_R J(J+1),
$$
independent of $M$. The $M$-degeneracy can be lifted if the molecule has a rotational magnetic moment $\frac{g \mu_N}{\hbar} \vec{J}$ with  $g$ the rotational g-factor and $\mu_N$ the nuclear magneton. In addition, a permanent or induced dipole moment $\vec{d}$ of the molecule can couple to any external electric field and also lead to an $M$-dependence of the rotational eigenstate energies.\\
\\
The energy contributions due to the rotational magnetic moment coupling to a nuclear magnetic moment $\frac{g_I \mu_N}{\hbar} \vec{I}$ ($\vec{I}$ is the nuclear spin operator and  $g_I$ the nuclear g-factor) and to an external static magnetic field $\vec{B}$ can be described by the Hamiltonian: 
\begin{equation}
    H^{(B)} = -\frac{g \mu_{N}}{\hbar} \; \vec{J}\cdot \vec{B}-\frac{g_{I} \mu_{N}}{\hbar} \; \vec{I}\cdot \vec{B} - \frac{2\pi c_{IJ}}{\hbar} \; \vec{I}\cdot \vec{J}
\end{equation}
where the $c_{IJ}$ is the spin rotation constant. The coupling of the magnetic moments to each other through $c_{IJ}$ competes with the coupling of each moment to the external magnetic field and the product states $\ket{J,M}\ket{I,M_I}$ are generally no longer energy eigenstates. In a coordinate frame where the static magnetic field is oriented along the $z$-direction, $\vec{B}=(0,0,B_z)$ the eigenstates of $H^{(B)}$ are also eigenstates of the sum of $z$-components of angular momenta $J_z+I_z$ with eigenvalues $m=M+M_I$. In the special case of $I=1/2$ that applies to CaH$^+$, the eigenstates can be found analytically in analogy to the Breit-Rabi solution for the angular momentum of a spin-1/2 electron coupled to a nuclear spin. The CaH$^+$ eigenstates can be labeled as $\ket{J,m,\pm}$ (see \cite{Chou2017} for details), but this approach does not apply for general $I>1/2$. However, it is worth noting that the extreme product states $\ket{J,\pm J}\ket{I,\pm I}$ where the rotation and the nuclear spin are fully aligned remain eigenstates of $H^{(B)}$, and  are only shifted in energy by the spin-rotation coupling and the interaction with the external magnetic field, but not coupled to each other or to any rotational states with $|M|\neq J$.\\
\\  
The interaction of the electric field due to the trap rf-drive with the permanent electric dipole moment $\vec{d}$ of the molecule is described by Eq. (1) in the main text,
\begin{equation}
    H = -\vec{d}\cdot \vec{E}_{\rm rf}\cos(\Omega_{\rm rf} t),
\end{equation}
with variables defined as stated in the main text. The field $\vec{E}_{\rm rf}$ does not couple to the nuclear spin but, within a certain $J$-manifold, it couples different product states than $H^{(B)}$. With respect to a coordinate system with quantization axis along $\vec{B}$, one can decompose $\vec{E}_{\rm rf}=(E_x,E_y,E_z)$ and $\vec{d}=(d_x,d_y,d_z)$ into their spherical tensor components
$$
\vec{E}_{\rm rf}= E_{\rm rf}(\epsilon_{-1} \vec{n}_{\sigma^-}+\epsilon_0 \vec{n}_\pi+\epsilon_{1} \vec{n}_{\sigma^+}), 
$$
$$
\vec{d}=d(\delta_{-1} \vec{n}_{\sigma^-}+\delta_0 \vec{n}_\pi+\delta_{1} \vec{n}_{\sigma^+}),
$$
with coefficients
$$
\epsilon_{-1}=\frac{E_x+i E_y}{\sqrt{2}E_{\rm rf}},~\epsilon_0=\frac{E_z}{E_{\rm rf}},~\epsilon_{1}=-\frac{E_x-i E_y}{\sqrt{2}E_{\rm rf}},~ E_{\rm rf}=|\vec{E}_{\rm rf}|,
$$
$$
\delta_{-1}=\frac{d_x+i d_y}{\sqrt{2}d},~\delta_0=\frac{d_z}{d},~\delta_{1}=-\frac{d_x-i d_y}{\sqrt{2}d},~ d=|\vec{d}|,
$$
and the orthonormal spherical basis vectors
$$
\vec{n}_{\sigma^-}=\frac{\vec{n}_x-i \vec{n}_y}{\sqrt{2}},~\vec{n}_\pi=\vec{n}_z,~\vec{n}_{\sigma^+}=-\frac{\vec{n}_x+i \vec{n}_y}{\sqrt{2}},
$$
written in terms of Cartesian unit vectors $\{\vec{n}_x,\vec{n}_y,\vec{n}_z\}$. With that, the matrix elements first order in $E_{\rm rf}$ is
$$
\bra{J',M'} H^{(1)} \ket{J,M}=-\bra{J',M'} \vec{d}\cdot \vec{E_{\rm rf}}\cos(\Omega_{\rm rf} t) \ket{J,M}= -d~ E_{\rm rf} \cos(\Omega_{\rm rf} t) \sum_{k=-1}^1 \bra{J',M'}\delta_k\Ket{J,M} \epsilon^*_k, 
$$
and only non-zero if $0\leq J =J' \pm 1$, $M'+k=M$ and $|M| \leq J$. In the interaction picture,  elements with $J=J'\pm 1$ rotate rapidly  (don't conserve energy), since $\hbar \Omega_{\rm rf} \ll |E_J -E_{J\pm 1}|$, where $E_J$ is the unperturbed rotational energy of state $\ket{J,M}$. This holds for most experimentally used rf drive frequencies and typical molecules and implies that there are negligible shifts or coupling due to $H$ at first order in $E_{\rm rf}$.\\
\\
The matrix elements second order in $E_{\rm rf}$ can be computed by considering that initial and final states are connected through intermediate states $\ket{J_i,M_i}$ and summing over all intermediate states for which products of two dipole matrix elements of the form
$$
\bra{J',M'}H \ket{J_i,M_i}\bra{J_i,M_i}H\ket{J,M} 
$$
do not vanish or rotate rapidly in the interaction picture. Since the products rotate at $|E_J-E_{J'}|/h$, slow rotation (approximate energy conservation) is only possible for $J=J'$. By adiabatically eliminating the intermediate levels, one arrives at the second order matrix elements stated in Eq. (2) of the main text,
\begin{eqnarray}
\bra{J',M'} H^{(2)} \ket{J,M}&=& \nonumber \\
\frac{\cos^2(\Omega_{\rm rf} t)}{4 h} \sum_{i} \frac{\bra{J',M'}\vec{d}\cdot\vec{E}_{\rm rf}\ket{i}\bra{i}\vec{d}\cdot\vec{E}_{\rm rf}\ket{J,M}}{f_{\rm rf}-\nu_{J,i}}
&+&\frac{\bra{J',M'}\vec{d}\cdot\vec{E}_{\rm rf}\ket{i}\bra{i}\vec{d}\cdot\vec{E}_{\rm rf}\ket{J,M}}{-f_{\rm rf}-\nu_{J,i}},\nonumber
\end{eqnarray}
where the sum runs over all intermediate states with non-zero matrix elements, namely $\ket{i}=\ket{J_i,M_i}$, $0\leq J_i=J\pm 1$, $M_i =M +\{-1,0,1\}$ and $|M_i| \leq J_i$. The frequency $\nu_{J,i} = (E_i-E_J)/h $ is proportional to the difference between rotational state energies $E_J$. We can also neglect the term rotating at $2\Omega_{\rm rf}$ in $\cos^2(\Omega_{\rm rf} t) =\frac{1}{2}+\frac{1}{2} \sin(2 \Omega_{\rm rf} t)$. Due to the dipole selection rules only second order couplings with  $|M-M'|\leq 2$ can be non-zero and for these cases the sums can be calculated explicitly and yield the  matrix elements stated in Eqns. (3) and (4) of the main text.\\
\\
Since the rf electric field does not interact with the nuclear spins, the matrix elements of product states can only be non-zero for identical nuclear spin states, 
$$
\bra{J,M}\bra{I,M_I}H^{(2)}\ket{I',M'_I} \ket{J,M'}= \bra{J,M} H^{(2)} \ket{J,M'}~\delta(I,I')~\delta(M_I,M'_I), 
$$
as indicated by the Kronecker-delta $\delta(i,j)$ for indices $i,j$. For general directions of $\vec{E}_{\rm rf}$, the matrix elements of $H^{(2)}$ and $H^{(B)}$ can couple a number of different product states and therefore finding the eigenstates and eigenvalues is not analytically straightforward. The total Hamiltonian $H^{(B)}+H^{(2)}$ for a certain rotational manifold with fixed $J$ can still be numerically diagonalized for general field directions and we have done this to estimate the effect of the rf electric field on the joint eigenstates of the $J=2$ manifold in the static magnetic and rf electric fields we used in our experiments (see below).\\
\\
%
%
When the directions of magnetic and rf electric field can be freely chosen, one can simplify the problem by aligning $\vec{E}_{\rm rf}$ parallel to $\vec{B}$. In this case, $\epsilon_0=1,~ \epsilon_{-1}=\epsilon_1 =0$ and
$$
\bra{J,M}\bra{I,M_I}H^{(2)}\ket{I,M_I} \ket{J,M'}= \bra{J,M} H^{(2)} \ket{J,M} \delta(M,M') 
$$
is diagonal in the product states. This implies that the extreme product states $\ket{J,\pm J}\ket{I,\pm I}$ are eigenstates of $H^{(B)}+H^{(2)}$ and are only shifted in energy but not mixed with other states. If the magnetic field is held constant and the magnitude of the parallel rf electric field is varied, one can determine the frequencies of transitions between extreme states, which should change linearly as a function of $E_{\rm rf}^2$. The slope of this linear relation can then be used to determine the permanent dipole moment of the molecule, while the intercept with $E_{\rm rf}^2=0$ is determined by the rotational spacing and the energy shifts of the extreme states due to $H^{(B)}$. The energy shift for the extreme states are the same for $M=\pm J$,
$$
\Delta E_{J}= -\frac{d^2 E_{\rm rf}^2}{4 h B_r(J+1)(2 J+3)},
$$
making transitions involving the lowest $J$ the most sensitive to $\vec{d}$ with the largest magnitude of shifts. For Raman-interactions $\Delta J=2$, and the extreme transitions $\ket{0,0}\ket{1/2,\pm 1/2} \rightarrow \ket{2,\pm 2}\ket{1/2,\pm 1/2}$ shift by
$$
\Delta \nu_{2,0}=\frac{\Delta E_0-\Delta E_2}{h} = -\frac{d^2 E_{\rm rf}^2}{12 h^2 B_r}\frac{6}{7}=\frac{H^{(2)}_{0,0}}{h}\frac{6}{7},
$$
while the electric dipole transitions $\ket{0,0}\ket{1/2,\pm 1/2} \rightarrow \ket{1,\pm 1}\ket{1/2,\pm 1/2}$ shift slightly less,
$$
\Delta \nu_{1,0} = -\frac{d^2 E_{\rm rf}^2}{12 h^2 B_r}\frac{7}{10}=\frac{H^{(2)}_{0,0}}{h}\frac{7}{10}.
$$
In practice one may also consider the resolution attainable on the transition, which could be favorable for the dipole transition, if the resolution is limited by the black-body lifetime of the involved rotational states. Finally, since the magnetic field shifts are equal and opposite for extreme levels with opposite signs, $m =\pm (J+1/2)$, while the electric field shifts are the same, slow magnetic field drifts can be averaged out by alternating frequency determinations between extreme level pairs. 
\section{Bounding the $J=2$ sublevel shifts}
\noindent By investigating several  molecular transitions within a given rotational manifold at a number of different rf electric field configurations (changing both magnitude and polarization) we can compare theoretical predictions of energy level shifts to what we see experimentally with the molecular ion. We measured two transition frequencies (and then can infer a third between two of the states) for both $J=1$ and $J=2$ under 5 different rf electric field configurations with differing direction and magnitude. These transitions ranged in frequency from 200~Hz to 17~kHz. Guided by previous theory values for $c_{IJ}$ and $g$ \cite{Chou2017} we explored the parameter space in that vicinity for the theoretical model, and find root-mean-square error of 86 Hz (73 Hz) for $J = 1$ (2) with respect to the experimentally determined transitions for the spin rotation constant $c_{I=1/2,J=1}= c_{I=1/2,J=2}\approx$ 8.06~kHz and g-factor $g\approx$ -1.39. These parameters are within 3$\%$ of their theoretically calculated values of 8.26~kHz and -1.35 respectively. This gives us confidence in our understanding of the molecular energy levels in the presence of an rf electric field.

In order to establish a bound on the shift of the $\ket{J=2,-3/2,-}$ state for the different experimental conditions of our rotational spectroscopy measurements, we use the model to examine the predicted energy shift for different rf electric fields. By doing so for a variety of $c_{IJ}$'s and $g$'s, we determine the typical behavior of the energy level when varying the rf electric field.  For rf electric fields we used for spectroscopy and with $c_{IJ}$'s and $g$'s varying $\pm$10$\%$ from the best fit values above, the energy of the $\ket{J=2,-3/2,-}$ state can vary and lead to a 0 to $-50$~Hz frequency shift from a zero rf field configuration to the largest rf electric field we used ($\sim$3100~V/m). We also note that the energy shift of this particular $J=2$ state is the least sensitive to rf electric field of any state in the $J=2$ rotational level manifold. When determining the slope of the $J'=2$ to $J''=0$ transition frequency with respect to electric field we assign a frequency uncertainty due to the shift of this $J=2$ state of 50~Hz standard deviation for each data point.

This uncertainty is added in quadrature with the transition frequency fit uncertainty in order to establish the total uncertainty in the transition frequency axis. Then by utilizing parametric bootstrapping as described above, we  perform a similar analysis to that used to establish the rf electric field-free transition frequency. The repeated fits to bootstrapped data points allow us to assign a confidence interval to the slope of the line of transition frequency vs  squared rf electric field magnitude, which is then converted to the electric dipole moment of the molecular ion.

\bibliographystyle{apsrev4-1}
\bibliography{micromotion}